\documentstyle[12pt,epsf]{article}

\begin{document}

\title{Self-Organized Criticality in the Olami-Feder-Christensen model}
\author{Josu\'{e} X. de Carvalho\footnote{E-mail:josue@if.usp.br} and Carmen P. C. Prado\footnote{E-mail:prado@if.usp.br} \\
Instituto de F\'{\i}sica\\
Universidade de S\~{a}o Paulo\\
Caixa Postal 66318\\
05315-970, S\~ao Paulo, SP, Brazil\\
}
\maketitle

\begin{abstract}
A system is in a self-organized critical state if the distribution of some
measured events obeys a power law. The finite-size
scaling of this distribution with the lattice size is usually
enough to assume that the system displays SOC. This approach, however, can be misleading. In
this work we analyze the behavior of the branching rate $\sigma $ of the
events to establish whether a system is in a critical state. We apply this
method to the Olami-Feder-Christensen model to obtain evidences that, in
contrast to previous results, the model is critical in the conservative
regime only.

PACs number(s): 64.60.L, 05.40, 05.70.L

\end{abstract}

In spite of many efforts and more than a decade of studies, the presence of
self-organized critical behavior in nature (and in some computer models) is
a matter of controversy. The concept of self-organized criticality (SOC) was
originally proposed by Bak, Tang and Wiesenfeld to describe the appearance of
scale invariance in nature. The idea was presented through the study of the
behavior of avalanches in a sandpile `toy' model \cite{BTW}. This simple
model displayed the fundamental properties associated with self-organized
criticality. Under a slow driven perturbation the system evolves to a
critical state, with no characteristic time and length scales. Once in this
state, the response of the system to the slow perturbation has no typical
length, and even a small perturbation (as the addition of a single grain of
sand) can start a big avalanche.

Avalanching behavior as well as scale invariance have been experimentally
observed in a variety of situations in nature, ranging from such different
phenomena as earthquakes\cite{gutemberg} or magnetic systems (the Barkhausen
effect)\cite{magnetic}, to biological problems such as evolution of species%
\cite{sneppen} or lung inflation\cite{lung}, just to give some examples.
Although by now the initial attempt to explain the appearance of all linear
scaling in nature through the concept of SOC may seem a little naive, the
ubiquity of its presence is still a strong suggestion that some kind of
`robust' and general mechanism may be behind many of these phenomena. The
concept of SOC has become polemic, and, up to now, there is no general
agreement about the ingredients necessary to create the self-organized
critical state. Particularly, there are discussions about the need of some
kind of local conservation as an essential ingredient of the system to
display SOC. The existence of SOC in non-conservative models would be highly
desirable in this context, since, in practice, some kind of dissipation is
always present in nature.

One of the best successful applications of the ideas of SOC for
non-conservative systems are the investigations of the
Olami-Feder-Christensen on a model for the dynamic of earthquakes (hereafter
called OFC model\cite{OFC}). In this model there is a parameter $\alpha $
that controls the level of conservation. Based on strong numerical evidences
\cite{evidencias} it has been taken as an example of a system that has
self-organized criticality in the non-conservative regime, that is, for $
\alpha <0.25$.

In this paper we revisited the OFC model, but with a different technique.
Instead of looking for power laws in distribution functions of avalanche
sizes versus lattice sizes, we looked at the behavior of the average
branching rate, both in the conservative and in the non-conservative regime.
In contrast to previous evidences, we concluded that the OFC model is
critical only for $\alpha \approx 0.25$ (that is, in the conservative
regime). For values of $\alpha $ close to but smaller than $0.25,$ this
model could be classified as `almost critical'. That means that, although
being critical only for $\alpha =0.25$, for all practical purposes the
system behaves as if it were critical for a wide range of values of $\alpha $
, with well-defined power laws over many decades.

In a recent paper, Kinouchi and Prado\cite{kino/prado} showed that some
models that exhibit numerical evidences of self-organized criticality in a
wide range of the coupling parameters were indeed what they called `almost
critical'. Through the analysis of the branching rate $\sigma $ as a
function of the dissipation parameter $\alpha ,$ they have shown that,
although those systems are critical only for $\alpha =\alpha _c$, there are
a rather large region around this point where approximate scale invariance
holds. They called this behavior `almost critical' since, in practice, it
can hardly be distinguished from `true' criticality based on the usual
numerical evidences only. By usual numerical evidences we mean power-law
behavior and scale invariance in distribution functions (the avalanche size
distribution function, for instance). They also suggested that the analysis
of the branching rate $\sigma $ (where $0\leq \sigma \leq 1$) as a function
of the coupling constant $\alpha $ could be a more efficient way to
determine whether a model is critical or not. To look for power-laws in
lattices of increasing sizes is not a very efficient way to determine if a
system is in fact critical, and this approach has already lead to mistakes%
\cite{canela}. If the analyzed lattices are not big enough, the distribution
functions of avalanche sizes $F(s)$ are power laws, even if the model does
not display SOC. Because the computational cost of simulating the OFC model
(and many others) in big lattices is prohibitive and there is no way to
know, beforehand, if the considered lattices are big enough to show the real
characteristics of the dynamical behavior of the system, such approach is
hardly conclusive.

It has been shown that some SOC models, with no spatial correlations and in
the thermodynamic limit, can be mapped into a branching process\cite{zapperi}%
. A branching process is a Markovian process and can be characterized by a
sequence of random variables $\left\{ P(n)\right\} _{n=0}^\infty $ , $n\in N$%
, where $P(n)$ represents the total number of individuals of the $n^{th}$
generation. Consider a group of individuals (ancestors) that can replicate,
giving birth to some descendants, and let $p_i$ $(i=0,1,...,\infty )$ be the
probability of an ancestor to give birth to $i$ descendants. Each of its
descendant in turn can give birth to other descendants with the same
probability $p_i$ so that $p_i$ does not depend on the previous generations
and on the number of descendants of other individuals in the same generation.
The branching rate, $\sigma =\sum\limits_{i=0}^\infty i\,p_i$, is then
defined as the average number of descendants per ancestor. It is a well known
result that, in order to have a critical branching process, one must have $%
\sigma =1$. Then the total number of descendants $P(n)$ in each generation
(the size of the `colony') behaves as a power law $P(n)\propto n^{-3/2}$%
\cite{harris} . On the basis of these considerations about the branching
rate, and using different approaches, several authors \cite{hakin} were able
to show that the random version of the OFC model was critical in the
conservative regime only.

Therefore, we decided to use this same approach to revisit the original
Olami-Feder-Christensen model\cite{OFC}. This coupled-map lattice model is
inspired on the spring block model developed by Burridge and Knopoff \cite
{burridge}. Each site $(\,i,\,j\,)$ of a square lattice is associated with a
continuous `energy' $F_{i\,j}$, initially set to a random value in the
interval $(0,F_c)$. The system is driven by a global perturbation that
increases the energy of all sites uniformly and simultaneously. This process
goes on until eventually one site becomes supercritical, that is, $%
F_{i\,j}\geq F_c$ . This site becomes unstable and the system then relaxes
according to the rules 
\begin{eqnarray*}
F_{i\,j} &\rightarrow &0,
\end{eqnarray*}
and
\begin{eqnarray*}
F_{n\,n} &\rightarrow &F_{n\,n}+\alpha \,F_{i\,j},
\end{eqnarray*}
where $F_{n\,n}$ are the four nearest neighbors of site $(\,i,\,j\,)$. The
parameter $\alpha $ controls the level of conservation of the model. If $%
\alpha =0.25$, the system is said to be `conservative', that is, all the
energy (or strength) lost by the site $(\,i,\,j\,)$ is distributed to its
neighbors. This relaxation rule can possibly produce a chain reaction that
only ends when all sites are stable again $(F_{i\,j}<F_c,\,\forall
\,i,j\,)$. As in the original work, we assume open boundaries. Also, as
shown in reference\cite{OFC}, one must have $\alpha <0.25$ to mimic the
dynamic of a real earthquake (some `energy' or `strength' is always lost
to the upper moving tectonic plate). This model is believed to display
self-organized criticality even when the dynamic is non-conservative $%
(0<\alpha <0.25)$. This is a result not yet fully understood, and it has
been a matter of controversy the value of the lower bound for $\alpha $ (if
it exists), under which the system has a localized behavior (note that we
know that $\alpha =0\Rightarrow \sigma =0,$ and, for $\alpha =0.25,$ we
should have $\sigma =1$). Because it is a model defined on a lattice,
analytical approaches are difficult and most of the results have been
obtained from computer simulations.

As the existence of a lattice introduces spatial correlations, it is not
possible to define the probability $p_i$ analytically. We estimate the
branching rate $\sigma $ numerically ($\sigma =<n_d>$, where $<n_d>$ is the
average number of supercritical sites (descendants) originated by an unstable
site). Just for comparison, we also study the random neighbor version of the
OFC model (R-OFC)\cite{R-OFC}, for which there are some analytical results%
\cite{hakin} showing that the model is critical for $\alpha =0.25$ only.

Our results are presented in Tables 1 and 2 and in Figures 1 to 3. We
checked the dependence of $\sigma $ on the lattice size (see Figures 2 and
3), and a special care has been taken to guarantee that the long transients
were eliminated. We also checked the effects of the boundaries. In the OFC model we
considered open boundaries to calculate $\sigma ,$ taking into account that
the average number of descendants for a boundary site is the number of
unstable sites it gives birth divided by the real number of neighbors of the
`ancestor' site (3 for a border site and 2 for a corner site). The R-OFC model was
simulated without borders. In most of
the cases, we first generated different stationary configurations from
different random initial configurations. The errors were estimated by
averaging results obtained for different initial configurations of the
lattice (the errors so obtained are usually bigger than the ones obtained by
averaging $\sigma $ during many generations, except when the system is
conservative). The number of iterations needed to reach the stationary state
is very big, and grows with the lattice size. In the OFC model the transient
is bigger for smaller values of $\alpha $, while in the R-OFC the transient
grows as $\alpha $ grows, making it impossible to simulate the case $\alpha
=0.25$ (the point in the graph in this case was obtained from theoretical
results).

\begin{table}
\begin{center}
\begin{tabular}{|c|c|c|c|c|c|c|}\hline
$\alpha$ & $\sigma$ & $\sigma_{b}$ & $<s>$& $L_{max}$  \\ \hline

 $0.15$ & $0.7052\pm 0.0002$  & $0.7151\pm 0.0002$ & $3.40\pm 0.02$ &$100$ \\
 \hline

 $0.18$ & $0.8361\pm 0.0003$  & $0.8430\pm 0.0003$ & $6.08\pm 0.08$ &$150$ \\
 \hline

 $0.21$ & $0.9125\pm 0.0002$  & $0.9205\pm 0.0002$ & $11.0\pm 0.6$ &$100$ \\
 \hline

 $0.22$ & $0.9546\pm 0.0009$ & $0.9581\pm 0.0009$ & $21.4\pm 0.4$ &$200$ \\
 \hline

 $0.23$ & $0.982\pm 0.001$  & $0.983\pm 0.001$ & $53\pm 3$ &$400$ \\
 \hline

 $0.24$ & $0.9938\pm 0.0004$  & $0.9946\pm 0.0004$ & $148\pm 9$ &$400$ \\
 \hline

 $0.25$ & $1.000003\pm 0.000009$  & $1.000068\pm 0.000009$ & $39839\pm 68$ 
&$400$ \\
 \hline

\end{tabular}
\caption{
 }
\end{center}
\end{table}

\begin{table}
\begin{center}
\begin{tabular}{|c|c|c|c|c|}\hline
$\alpha$ & $\sigma$  & $<s>$& $L_{max}$  \\ \hline

 $0.15$ & $0.6006\pm 0.0003$  & $2.083\pm 0.001$ &$100$ \\
 \hline

 $0.18$ & $0.7140\pm 0.0003$  & $3.497\pm 0.004$ &$100$ \\
 \hline

 $0.21$ & $0.8595\pm 0.0002$  & $7.12\pm 0.01$ & $400$ \\
 \hline

 $0.22$ & $0.9297\pm 0.0002$  & $14.22\pm 0.04$ &$500$ \\
 \hline

 $0.23$ & $0.9876\pm 0.0002$  &  $81\pm 1$ &$800$ \\
 \hline

 $0.24$ & $0.99923\pm 0.00008$  &  $1306\pm 80$ &$1000$ \\
 \hline

\end{tabular}
\caption{}
\end{center}
\end{table}

Once we were sure to have a stationary configuration, we analyzed the
statistics of $100\,000$ to $5\,000\,000$ avalanches in the stationary
state, to obtain (a) the average avalanche size $<s>$, (b) the branching
rate $\sigma $ (weighting border sites), (c) the branching rate in the bulk $%
\sigma _b$ (taking into account only sites in the bulk), and (d) the average
number of generations in an avalanching process $<n>$. Table 1 shows the
results for the OFC model and Table 2 shows the results for the R-OFC model.

There are no relevant differences between the behaviors of the OFC and the
R-OFC models. For both of them, $\sigma (\alpha )\rightarrow 1$ smoothly
from below as $\alpha \rightarrow 0.25$ , with no sign of any kind of
discontinuity in its behavior. Also, as can be seen in Figure 1, $\sigma
_{OFC}<\sigma _{R-OFC},$ for $0.22\leq \alpha <0.25$. From theoretical
considerations\cite{hakin}, we know that $\sigma _{R-OFC}<1$ for $\alpha
<0.25$.

In Figures 2 and 3 we present the dependence of $\sigma $ on the lattice
size for the OFC and the R-OFC models. These figures show that $\sigma $
grows almost linearly with $1/L$ with no suggestion that $\sigma \rightarrow 1$
as $1/L\rightarrow 0$. The behavior of the system seems to be qualitatively
different only if $\alpha =0.25$ (conservative case).

We also checked the dependence of $\sigma $ on the generation $n$ within
an avalanching process. We see that $\sigma (n)$  converges relatively fast
to an asymptotic value \cite{futuro}. None of our
conclusions were affected if we considered these asymptotic values of $%
\sigma (n)$ instead of the average value.

The existence of SOC in the non-conservative regime of the OFC model has
been accepted based mainly on numerical results of a work done by Middleton
and Tang \cite{evidencias} in 1995. In this paper, the authors showed how
the natural tendency of this model to synchronize is destroyed by
inhomogeneities introduced by the asymmetries of the boundaries, creating
long-range correlations and leading to a power-law behavior in the
distribution of avalanche sizes. The apparent contradiction between this
result and ours can be understood from the conclusions of Kinouchi and Prado 
\cite{kino/prado}. In this paper, the study of two different models with an
analytical solution (the extremal Feder and Feder model, EFF, with and
without noise), shows that the effect of noise is to enlarge the region
where the system displays an apparent critical behavior, leading to what was
called `almost criticality'. The EFF model with noise displays a power law
behavior (although it is not critical). In contrast, in the noiseless model,
large avalanches occur in the conservative limit only. This also seems to be
the case of the OFC model. The randomness introduced by the asymmetries of
the boundaries creates correlations that enlarge the critical region leading
to an `almost critical' behavior, although it is not enough to ensure true
criticality.

In conclusion, we showed that the analysis of $\sigma (\alpha )$ is a
complementary approach to define if a model is or is not critical. This new
method revealed that the behavior of the OFC model is qualitatively
identical to the behavior of the R-OFC. In contrast to previous results, the
Olami-Feder-Christensen model seems to be critical only in the conservative
regime, that is for $\alpha =0.25$. Both models are `almost' critical in
the sense defined in reference \cite{kino/prado}: $\sigma \approx 1$ when $%
\alpha \approx 0.25$, leading to a power law behavior of the avalanche sizes
for many decades, and making it (almost) impossible to distinguish this
behavior from `true' self-organized criticality based on the observation of
power-laws and finite-size scaling fits.

The authors acknowledge Dr. Osame Kinouchi for helpful discussions and
suggestions. J. X. Carvalho acknowledges the Brazilian agency CAPES for
financial support.

\newpage
{\Large \textbf{FIGURE CAPTIONS}}

\textbf{Table 1:}Values of $\sigma $, $\sigma _b$ and $<s>$ for different
values of the conservative parameter $\alpha $ in the
Olami-Feder-Christensen (OFC) model. Results presented are those obtained
with the biggest lattice  ($L_{max}$)  we were able to simulate. They 
represent the average of results obtained for different initial
configurations and the errors are the errors associated with those averages.

\textbf{Table 2:}Values of $\sigma $, $\sigma _b$ and $<s>$ for different
values of the conservative parameter $\alpha $ in the random version of the
Olami-Feder-Christensen (R-OFC) model. Results presented are those obtained
with  the biggest lattice  ($L_{max}$)  we were able to simulate. They
represent the average of results obtained for different initial
configurations and the errors are the errors associated with those averages.

\textbf{Figure 1:} Branching rate as a function of the conservation
parameter $\alpha $. Squares  refer to the Olami-Feder-Christensen model
(OFC)  and circles to the Random version of the OFC model (R-OFC). In all
cases the lattice size is $L=100.$

\textbf{Figure 2:} Branching rate as a function of the inverse of lattice
size $(1/L)$ for the Random version of Olami-Feder-Christensen model.
Different curves refer to different levels of conservation ($\alpha
=0.22,\,0.23$ and $0.24\,$). We can see that even for $\alpha =0.24,$ if we
let $L\rightarrow \infty $, the branching rate $\sigma $ tends to a value
smaller than $1.$ The system shows a qualitatively different behavior only
if $\alpha =0.25$.

\textbf{Figure 3:} Branching rate as a function of the inverse of lattice
size $(1/L)$ for the  Olami-Feder-Christensen model. Different curves refer
to different levels of conservation ($\alpha =0.23,\,0.24$ and $0.25\,$). We
can see that even for $\alpha =0.24,$ if we let $L\rightarrow \infty $, the
branching rate $\sigma $ tends to a value smaller than $1.$ Note that in the
conservative case ($\alpha =0.25$) $\sigma $ is almost 1.00 even to very
small lattices.


\begin{thebibliography}{99}
\bibitem{futuro} The evolution of $\sigma$ within an avalanching process is
the current subject of another work and will be published elswhere. We called
each time step in an avalanching process a "generation". Preliminary results
indicate that $\sigma(n)$ converges relatively fast (10 generations are enough) to 
an assymptotic value sligthly smaller than the average use in this paper,
 $\sigma =<n_d>$. In the OFC model, for instance, we obtained, for $\alpha = 0.25$, 
$\sigma_{\infty}=0.99787\pm 0.00006$, and for $\alpha = 0.24$, 
$\sigma_{\infty}=0.9881\pm 0.0006$.
\bibitem{R-OFC}  The R-OFC model follows the same dynamical rules of the OFC
model, except for the way energy is distributed among the sites. When a site 
$(i,j)$ becomes supercritical, instead of spreading its energy to the (four)
nearest neighbors ($F_{nn}\rightarrow F_{nn}+\alpha \,F_{ij}$), it spreads
energy to sites chosen at random. Each time a site relaxes, new `random
neighbors' are chosen, in a process that has been called annealed
randomness in literature. Such random versions have been constructed and
studied because they can be thought as numerical simulations of mean-field
approaches of lattice models. For more details about the R-OFC model, see,
for instance, the articles of reference\cite{hakin} below.
\end{thebibliography}
\end{document}